# The Fractional Schrödinger Equation with the Generalized Woods-Saxon Potential

M. Abu-Shady[1]
Department of Mathematics and computer science, Faculty of Science, Menoufia University, Egypt[1].
Etido P. Inyang[2]
Department of Physics, National Open University of Nigeria, Jabi, Abuja, Nigeria[2]

The bound state energy eigenvalues and the corresponding eigenfunctions of the generalized Woods-Saxon potential reported in [Phys. Rev. C 72, 027001 (2005)] is extended to the fractional forms using the generalized fractional derivative and the fractional Nikiforov-Uvarov (NU) technique. Analytical solutions of bound states of the Schrodinger equation for the present potential are obtained in the terms of fractional Jacobi polynomials. It is demonstrated that the classical results are a special case of the present results at $\alpha = \beta = 1$. Therefore, the present results play important role in molecular chemistry and nuclear physics.
**Keywords:** generalized Fractional derivative, Schrödinger equation, Nikiforov-Uvarov method

## 1-Introduction

The fractional calculus (FC) has numerous applications in all connected fields of science and engineering [1]. However, the use of this extremely powerful tool in many studies is still in its infancy. Fractional calculus has recently expanded its scope to include the dynamics of the complicated real world, and new concepts are now being put to the test on actual data. Although FC has been around for a while and is used in numerous scientific and technical sectors, the FC still has a crucial role to play in promoting applications. Many theoretical FC researchers are likewise unfamiliar with the application-related aspects. Because FC is not universal and has specific applications, we should understand this and provide examples of some significant FC applications that have been successful in the past to serve as a model for future FC application research. The FC has grown throughout decades in numerous disciplines of mathematics, but until recently they had little use in physics and other mathematically focused sciences.

There are now more and more physics study fields that use FC, which indicates that the situation is starting to change [2, 3]. Scholars in physics and its related fields have recently become interested in the applications of FC to the Schrodinger equation (SE). For instance, Laskin [4] explored the fractional SE that contains the quantum Riesz fractional operator and the Caputo fractional derivative (FD). In order to solve the local fractional SE for the harmonic oscillator potential, the Hulthen potential (HP), and the Woods-Saxon potential (WSP), Karayer et al. [5] deduced the conformable fractional form of the NU technique. By utilizing the fractional version of the NU technique, Karayer et al. [6] have investigated the analytical solutions of the local Klein-Gordon problem for the generalized HP. The applications of FC in complicated and nonlinear physics were also presented by Baleanu et al. [7]. Another development was the study in [9-12] of the energy spectrum of heavy quarkonium in the context of fractional SE with an extended Cornell potential model in different systems. To examine the fractional version of Newtonian mechanics, conformable FD and integral have been used by Chung [13]. The fractional parameter (FP) $0 < \alpha \leq 1$ is connected to the space-roughness time's properties through the FC and its use in quantum physics. Additionally, the nature of wave equation solutions for different values of the FP indicates the fundamental behavior of the quantum mechanical systems [14].

Abu-Shady and Kaabar, recently introduced the generalized fractional derivative in [15, 16] that gives advantageous results more than the classical definitions. In addition, the definition gives good results in applying to different models such as in Refs. [17-19].

The WSP is a short-range potential and is used to study the nuclear structure within the shell model [20]. This potential has been presented in many forms to investigate the elastic and quasi-elastic scattering of nuclear particles. The usual (q = 1) and the q-deformed WS potentials have been applied in nuclear calculations [21]. The helium model and the nonlinear scalar theory of mesons both use it to explore the behavior of valence electrons in metallic systems. [22].

The WSP and its various modifications have been crucial in microscopic physics in determining the energy level spacing, particle number dependence of energy quantities, and universal properties of electron distributions in atoms, nuclei, and atomic clusters because they can be used to describe the interaction of a neutron with a single heavy-ion nucleus as well as for the optical potential model [23]. We are motivated to consider the solutions of the fractional SE for the generalized WSP using the generalized fractional (GF) NU method. This work is generalized to the work reported in [24] in the fractional model.

The following is how the paper is set up: The GF-NU approach is briefly presented in Section 2. In Section 3, the GF-NU technique is applied on the fractional Schrödinger equation. The results are discussed in Section 4. The overview and conclusion are offered in Section 5.

## 2-The Generalized Fractional NU method

This section provides a brief explanation of the GF-NU technique for solving the generalized fractional differential equation that has the following equation (see Refs. [6, 15] for more information).

$$D^\alpha \left[ D^\alpha \Psi(s) \right] + \frac{\bar{\tau}(s)}{\sigma(s)} D^\alpha \Psi(s) + \frac{\tilde{\sigma}(s)}{\sigma^2(s)} \Psi(s) = 0, \qquad (1)$$

where $\sigma(s)$ and $\tilde{\sigma}(s)$ are polynomials of maximum second degree of $\alpha$ and $2\alpha$, respectively, and $\bar{\tau}(s)$ has a maximum degree of $\alpha$
where
$$D^\alpha \Psi(s) = I s^{1-\alpha} \Psi'(s), \qquad (2)$$
$$D^\alpha \left[ D^\alpha \Psi(s) \right] = I^2 \left[ (1-\alpha) s^{1-2\alpha} \Psi'(s) + s^{2-2\alpha} \Psi''(s) \right]. \qquad (3)$$
where
$$I = \frac{\Gamma(\beta)}{\Gamma(\beta - \alpha + 1)} \qquad (4)$$

where $0 < \alpha \leq 1$ and $0 < \beta \leq 1$. Substituting by Eqs. (2) and (3) into Eq. (1), we obtain

$$\Psi''(s) + \frac{\bar{\tau}_f(s)}{\sigma_f(s)} \Psi'(s) + \frac{\tilde{\sigma}_f(s)}{\sigma_f^2(s)} \Psi(s) = 0, \qquad (5)$$

where,
$$\bar{\tau}_f(s) = (1-\alpha) s^{-\alpha} \sigma(s) + I^{-2} \bar{\tau}(s), \sigma_f(s) = s^{1-\alpha} \sigma(s), \tilde{\sigma}_f(s) = I^{-2} \tilde{\sigma}(s). \qquad (6)$$

If one works with the transformation, one may use the separation of variables to determine the specific solution of Eq.(5).
$$\Psi(s) = \Phi(s) \chi(s), \qquad (7)$$
it is reduced to the following hypergeometric equation.

$$\sigma_f(s)\chi''(s) + \tau_f(s)\chi'(s) + \lambda\chi(s) = 0, \tag{8}$$

where

$$\sigma_f(s) = \pi_f(s)\frac{\Phi(s)}{\Phi'(s)}, \tag{9}$$

$$\tau_f(s) = \bar{\tau}_f(s) + 2\pi_f(s); \quad \tau'_f(s) < 0, \tag{10}$$

and

$$\lambda = \lambda_n = -n\tau'_f(s) - \frac{n(n-1)}{2}\sigma''_f(s), n = 0,1,2,... \tag{11}$$

$\chi(s) = \chi_n(s)$ It has the following form and is an n-degree polynomial that satisfies the hypergeometric equation.

$$\chi_n(s) = \frac{B_n}{\rho_n}\frac{d^n}{ds^n}(\sigma''_f(s)\rho(s)), \tag{12}$$

where $B_n$ is a normalization constant and $\rho(s)$ is a weight function which satisfies the following equation

$$\frac{d}{ds}\omega(s) = \frac{\tau(s)}{\sigma_f(s)}\omega(s); \quad \omega(s) = \sigma_f(s)\rho(s), \tag{13}$$

$$\pi_f(s) = \frac{\sigma'_f(s) - \bar{\tau}_f(s)}{2} \pm \sqrt{(\frac{\sigma'_f(s) - \bar{\tau}_f(s)}{2})^2 - \tilde{\sigma}_f(s) + K\sigma_f(s)}, \tag{14}$$

and

$$\lambda = K + \pi'_f(s), \tag{15}$$

the $\pi_f(s)$ is a first-degree polynomial. If the expressions beneath the square root are squares of expressions, it is feasible to determine the values of K in Eq. (14). If its discriminate is zero.

## 3-The Generalized Fractional of Schrödinger Equation

The generalized WSP takes the form [24]

$$V(r) = -\frac{V_0}{1 + qe^{2\beta_1 r}} - \frac{ce^{2\beta_1 r}}{\left(1 + qe^{2\beta_1 r}\right)^2} \tag{16}$$

where $V_0$ is the potential depth, q is a real parameter, and $c$ is the surface thickness. This is often modified to reflect the experimental ionization energy values.
By substituting by Eq. (14), we can write Schrodinger equation [24]

$$[\frac{d^2}{dr^2}+\frac{2\mu}{\hbar^2}(E+\frac{V_0}{1+qe^{2\beta_1 r}}+\frac{ce^{2\beta_1 r}}{(1+qe^{2\beta_1 r})^2})]R(r)=0. \quad (17)$$

By assuming $x=-e^{2\beta_1 r}$, Eq. (17) takes the following form

$$[\frac{d^2}{dr^2}+\frac{2\mu}{\hbar^2}(E+\frac{V_0}{1+qe^{2\beta_1 r}}+\frac{ce^{2\beta_1 r}}{(1+qe^{2\beta_1 r})^2})]R(r)=0. \quad (18)$$

We introduce the following dimensional parameters:
where

$$\epsilon=-\frac{\mu E}{2\hbar^2 \beta_1^2}, \beta=\frac{mV_0}{2\hbar^2 \beta_1^2}, \gamma=\frac{\mu C}{2\hbar^2 \beta_1^2} \quad (19)$$

The following equation is obtained

$$[\frac{d^2}{dx^2}+\frac{1-qx}{x(1-qx)}\frac{d}{dx}+\frac{1}{s^2(1-qs)^2}(-\varepsilon q^2 x^2+(2\varepsilon q-\beta q-\gamma)x+\beta-\varepsilon)]R(x)=0. \quad (20)$$

To transfer Eq. (20) to the fractional form as in Eq. (1)

$$[D^\alpha[D^\alpha R(x)]+\frac{1-qx^\alpha}{x^\alpha(1-qx^\alpha)}D^\alpha R(x)+$$

$$\frac{1}{x^{2\alpha}(1-qx^\alpha)^2}(-\varepsilon q^2 x^{2\alpha}+(2\varepsilon q-\beta q-\gamma)x^\alpha+\beta-\varepsilon)]R(x)=0.$$

(21)

Substituting by Eqs. (2) and (3) into (21), we obtain

$$R''(x)+\frac{\bar{\tau}_f(x)}{\sigma_f(x)}R'(x)+\frac{\tilde{\sigma}_f(x)}{\sigma_f^2(x)}R(x)=0, \quad (22)$$

where

$$\bar{\tau}_f(s)=(1-\alpha)(1-qx^\alpha)+I^{-2}(1-qx^\alpha), \quad (23)$$

$$\sigma_f(s)=x(1-qx^\alpha), \quad (24)$$

$$\tilde{\sigma}_f(s)=I^{-2}(-\varepsilon q^2 x^{2\alpha}+(2\varepsilon q-\beta q-\gamma)x^\alpha+\beta-\varepsilon). \quad (25)$$

Using Eq. (14)

$$\pi_f = \frac{\left(-2+I^{-2}\right)qx^\alpha + \alpha - I^{-2}}{2} \pm \frac{1}{2}\sqrt{(A_1 - 4qw)x^{2\alpha} + 4(A_2 + w)x^\alpha + A_3} \qquad (26)$$

where

$$A_1 = \left(-2+I^{-2}\right)q^2 + 4I^{-2}\varepsilon q^2 \qquad (27)$$

$$A_2 = \frac{1}{2}\left(-2+I^{-2}\right)\left(\alpha - I^{-2}\right)q - I^{-2}\left(2\varepsilon q - \beta q - \gamma\right) \qquad (28)$$

$$A_3 = 4\left(\alpha - I^{-2}\right)q + 4(\varepsilon - \beta) \qquad (29)$$

The constant $w$ is selected so that the discriminant of the function under the square roots equals zero, giving the function a double zero. Hence, In addition, $k = wx^{\alpha-1}$ that defined in the following equation

$$w_\pm = \frac{-(8A_2 + 4A_3 q) \pm \sqrt{(8A_2 + 4A_3 q)^2 - 16(A_2^2 - 4A_1 A_3)}}{8} \qquad (30)$$

So, we can write Eq. (26) as follows

$$\pi_f = \frac{\left(-2+I^{-2}\right)qx^\alpha + \alpha - I^{-2}}{2} \pm \frac{1}{2}\left[\begin{array}{l}\sqrt{(A_1 - 4qw_+)}x^\alpha + 2\sqrt{(\varepsilon - \beta) + \left(\alpha - I^{-2}\right)^2} \\ \sqrt{(A_1 - 4qw_-)}x^\alpha - 2\sqrt{(\varepsilon - \beta) + \left(\alpha - I^{-2}\right)^2}\end{array}\right] \qquad (31)$$

By using Eq. (8), we write and select a negative sign as in Ref. [24]

$$\tau_f(s) = (1-\alpha)(1-qx^\alpha) + I^{-2}(1-qx^\alpha) + \left(-2+I^{-2}\right)qx^\alpha + \alpha - I^{-2}$$
$$- \left[\sqrt{(A_1 - 4qw_-)}x^\alpha - 2\sqrt{(\varepsilon - \beta) + \left(\alpha - I^{-2}\right)^2}\right] \qquad (32)$$

using Eqs. 11 and 15, we can write

$$\lambda_n = n(1-\alpha)\alpha q x^{\alpha-1} + nI^{-2}\alpha q x^{\alpha-1} - n\left(-2+I^{-2}\right)\alpha q x^{\alpha-1} + \alpha n\sqrt{(A_1 - 4qw_-)}x^{\alpha-1}$$
$$+ \frac{n(n-1)}{2}(1+\alpha)\alpha q x^{\alpha-1}$$

$$(33)$$

$$\lambda = w_- x^{\alpha-1} + \frac{1}{2}\left(-2+I^{-2}\right)\alpha q x^{\alpha-1} - \frac{1}{2}\left[\alpha\sqrt{(A_1 - 4qw_-)}x^{\alpha-1}\right], \qquad (34)$$

by using the $\lambda_n = \lambda,$ we obtain the energy eigenvalue in the fractional form

$$w_- + \frac{1}{2}(-2+I^{-2})\alpha q - \frac{1}{2}\left[\alpha\sqrt{(A_1-4qw_-)}x^{\alpha-1}\right]$$
$$= n(1-\alpha)\alpha q + nI^{-2}\alpha q - n(-2+I^{-2})q + \alpha n\sqrt{(A_1-4qw_-)} + \frac{n(n-1)}{2}(1+\alpha)\alpha q. \tag{35}$$

**The special case at** $\alpha = \beta = 1, \Rightarrow I = 1$

$$\varepsilon_{nq} = \frac{1}{16}\left[\sqrt{1+\frac{4\gamma}{q}}+(1+2n)\right]^2 + \frac{\beta^2}{\left[\sqrt{1+\frac{4\gamma}{q}}+(1+2n)\right]^2} + \frac{\beta}{2} \tag{36}$$

Eq. (36) is compatible with Ref. [24].

Let us now find the corresponding eigenfunctions as in Ref. [24]. It is necessary to identify the hypergeometric function that solves the differential equation in order to determine the polynomial solutions of the hypergeometric function $\rho(x)$ satisfying the equation $\left[\sigma_f \rho\right]' = \tau_f \sigma_f$. Thus, $\rho(x)$ is calculated as

$$\rho(x) = \frac{x^{A_{11}}}{(1-qx^\alpha)^{\frac{A_{11}q+B_{11}}{\alpha q}}} \tag{37}$$

By using the following relation

$$y_n(x) = \frac{B_n}{\rho(x)}\frac{d^n}{dx^n}\left(\sigma_f(x)^n \rho(x)\right) \tag{38}$$

where $B_n$ is a normalization constant, we obtain

$$y_n(x) = \frac{B_n}{x^{A_{11}}}(1-qx^\alpha)^{-\frac{A_{11}q+B_{11}}{\alpha q}}\frac{d^n}{dx^n}\left(x^{n+A_{11}}(1-qx^\alpha)^{n-\frac{A_{11}q+B_{11}}{\alpha q}}\right) \tag{39}$$

where

$$A_{11} = 2\sqrt{(\varepsilon-\beta)+(\alpha-I^{-2})^2} \tag{40}$$

$$B_{11} = 2\alpha q - I^{-2}q + (-2+I^{-2})q - \sqrt{(A_1-4qw)} \tag{41}$$

By using the relation $\frac{\Phi'(x)}{\Phi(x)} = \frac{\pi_f(x)}{\sigma_f(x)}$, we can obtain

$$\Phi(x) = \frac{x^c}{(1-qx^\alpha)^{\frac{Cq+D}{\alpha q}}} \tag{42}$$

where

$$C = \frac{1}{2}(\alpha - I^{-2}) + \sqrt{(\varepsilon - \beta) + (\alpha - I^{-2})^2} \tag{43}$$

$$D = \frac{1}{2}(-2 + I^{-2})q - \frac{1}{2}\sqrt{(A_1 - 4qw)} \tag{44}$$

Thus, we can write final form the corresponding wave function $R(x) = y_n(x)\Phi(x)$ as follows

$$R(x) = A_n x^{-(C+A_{11})} (1-qx^\alpha)^{-\frac{A_{11}q+B_{11}+Cq+D}{\alpha q}} \frac{d^n}{dx^n}\left( x^{n+A_{11}} (1-qx^\alpha)^{n-\frac{A_{11}q+B_{11}}{\alpha q}} \right)$$

$$= A_n x^{-(C+A_{11})} (1-qx^\alpha)^{-\frac{A_{11}q+B_{11}+Cq+D}{\alpha q}} P_n^{\left(A_{11}, \frac{A_{11}q+B_{11}}{\alpha q}\right)} (1-qx^\alpha) \tag{45}$$

where $A_n$ is the normalization constant and $P_n^{\left(A_{11}, \frac{A_{11}q+B_{11}}{\alpha q}\right)}$ is the orthogonal Jacobi polynomials. At $\alpha = \beta = 1$ and $q = 1$, we obtain the special of classical case with compatible with Ref. [24].

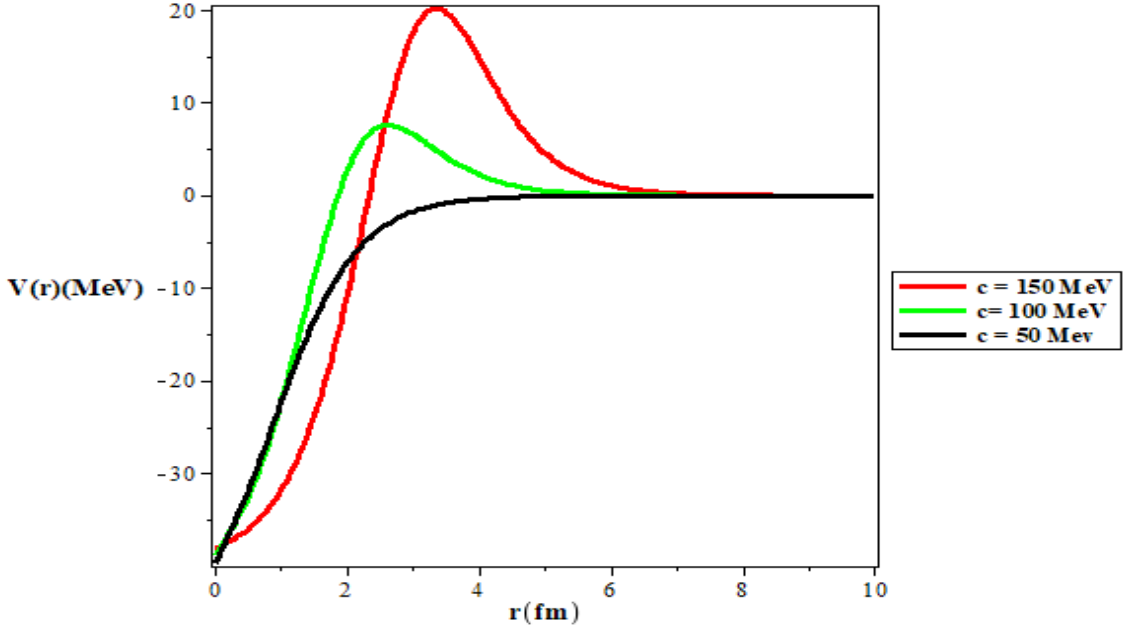

Fig. 1. Variation of the generalized Woods-Saxon potential as a function of *r*.

In Fig. 1, the variation of the generalized WSP is plotted where the empirical values are taken from Perey *et al.* [25] as $r_0 = 1.285$ fm and $a = 0.65$ fm. Moreover, the WSP parameter is investigated at $V0 \approx 40.5 + 0.13 A$ MeV.

Here, *A* is the atomic mass number of target nucleus. We note that the potential shifts to higher values by increasing parameter *c*.

## 4-Summary and Conclusion

We have adopted a generalized WSP to obtain the solutions of the fractional SE using the GF-NU method. Analytical solutions are obtained for the eigenvalues and eigenfunctions in the fractional forms. The results of Ref. [24] are obtained as a special cases at $\alpha = \beta = 1$. The present results are not considered in recent work. Therefore, the present results play an important role in molecular physics and nuclear physics. We hope to extend this work to hot and dense media and/or the present of magnetic field as future works as in Refs. [26-29].

## 5-References